%% file: main.tex
\documentclass[twocolumn]{aastex631}

\shorttitle{Limits on Leptonic TeV Emission from the Cygnus Cocoon with {\it Swift}-XRT}
\shortauthors{Guevel et al.}
\graphicspath{{./}{figures/}}

\begin{document}

\title{Limits on Leptonic TeV Emission from the Cygnus Cocoon with {\it Swift}-XRT}

\correspondingauthor{David Guevel}
\email{guevel@wisc.edu}

\author[0000-0002-0870-2328]{David Guevel}
\affiliation{Department of Physics, Wisconsin IceCube Particle Astrophysics Center, University of Wisconsin, Madison, WI, USA}

\author[0000-0001-5186-5950]{Andrew Beardmore}
\affiliation{School of Physics \& Astronomy, University of Leicester, 
University Road,  
Leicester, LE1 7RH, United Kingdom}

\author[0000-0001-5624-2613]{Kim L. Page}
\affiliation{School of Physics \& Astronomy, University of Leicester, 
University Road,  
Leicester, LE1 7RH, United Kingdom}

\author{Amy Lien}
\affiliation{University of Tampa, 
401 W Kennedy Blvd,  
Tampa, FL, USA}

\author[0000-0002-5387-8138]{Ke Fang}
\affiliation{Department of Physics, Wisconsin IceCube Particle Astrophysics Center, University of Wisconsin, Madison, WI, USA}

\author[0000-0001-7523-570X]{Luigi Tibaldo}
\affiliation{IRAP, Université de Toulouse, CNRS, CNES, UPS, 9 avenue Colonel Roche, 31028 Toulouse, Cedex 4, France}

\author{Sabrina Casanova}
\affiliation{Institute of Nuclear Physics Polish Academy of Sciences, PL-31342 IFJ-PAN, Krakow, Poland}

\author{Petra Huentemeyer}
\affiliation{Department of Physics, Michigan Technological University, Houghton, MI, USA}

\begin{abstract}
$\gamma$-ray observations of the Cygnus Cocoon, an extended source surrounding the Cygnus X star-forming region, suggest the presence of a cosmic ray accelerator reaching energies up to a few PeV.
The very-high-energy (VHE; 0.1-100~TeV) $\gamma$-ray emission may be explained by the interaction of cosmic-ray hadrons with matter inside the Cocoon, but an origin of inverse Compton radiation by relativistic electrons cannot be ruled out.
Inverse Compton $\gamma$-rays at VHE are accompanied by synchrotron radiation peaked in X-rays. Hence,
X-ray observations may probe the electron population and magnetic field of the source.
We observed eleven fields in or near the Cygnus Cocoon with the Neil Gehrels Swift Observatory's X-Ray Telescope ({\it Swift}-XRT) totaling 110 ksec. 
We fit the fields to a Galactic and extra-galactic background model and performed a log-likelihood ratio test for an additional diffuse component.
We found no significant additional emission and established upper limits in each field.
By assuming that the X-ray intensity traces the TeV intensity and follows an $dN/dE\propto E^{-2.5}$ spectrum, we obtained a 90\% upper limit of $F_X < 8.7\times 10^{-11}\rm~erg\,cm^{-2}\,s^{-1}$ or $< 5.2\times 10^{-11}\rm~erg\,cm^{-2}\,s^{-1}$ on the X-ray flux of the entire Cygnus Cocoon between 2 and 10~keV depending on the choice of hydrogen column density model.
This suggests that no more than one quarter of the $\gamma$-ray flux at 1~TeV is produced by inverse Compton scattering, when assuming an equipartition magnetic field of $\sim 20\,\mu$G.
\end{abstract}


\section{Introduction} \label{sec:intro}

The cosmic ray spectral energy distribution features a spectral break, commonly called ``the knee," at a few PeV.
The knee is believed to mark the transition from Galactic to extra-galactic cosmic ray origin.
The accelerators that produce the highest energy Galactic cosmic rays, known as PeVatrons, have not been conclusively identified.
The Cygnus Cocoon, from which $\gamma$-rays above 100~TeV have been detected \citep{abeysekaraHAWCObservationsAcceleration2021,tibetasgcollaborationGammaRayObservationCygnus2021, caoUltrahighenergyPhotonsPetaelectronvolts2021}, is a prime PeVatron candidate.
The Cygnus Cocoon contains the Cyg OB2 stellar association and NGC 6910 stellar cluster, as well as $\gamma$-Cygni Supernova Remnant but extends beyond them.
The winds of massive stellar clusters inside the Cocoon may potentially accelerate particles to PeV energies  \citep{morlinoParticleAccelerationWinds2021, vieuCosmicRayProduction2022, vieuCanSuperbubblesAccelerate2022}.

GeV and TeV $\gamma$-rays have been seen by the {\it Fermi} Large Area Telescope (LAT;  \citealp{atwoodLARGEAREATELESCOPE2009}) up to 100 GeV, the High Altitude Water Cherenkov Observatory (HAWC; \citealp{abeysekaraSensitivityHighAltitude2013}) beyond 100 TeV \citep{ackermannCocoonFreshlyAccelerated2011, abeysekaraHAWCObservationsAcceleration2021}, and by the Large High Altitude Air Shower Observatory (LHAASO) up to 1.4 PeV \citep{caoUltrahighenergyPhotonsPetaelectronvolts2021}.
The GeV emission detected by the {\it Fermi}-LAT indicates an cosmic-ray energy density that is 50\% larger than the ISM with a harder spectrum \citep{ackermannCocoonFreshlyAccelerated2011}, suggesting that cosmic ray particles are freshly accelerated at the Cocoon. The GeV emission may be explained by either leptonic or hadronic models.
The HAWC Collaboration found a change in the spectral shape at around 1 TeV and suggests that the morphology and spectral shape are consistent with proton interaction, though a leptonic contribution cannot be ruled out \citep{abeysekaraHAWCObservationsAcceleration2021}.

The electron population that inverse Compton scatters will necessarily produce synchrotron X-rays in the presence of a magnetic field.
The Klein Nishina effect suppresses inverse Compton scattering for optical and UV photons so that the TeV emission mostly traces the diffuse dust radiation field rather than the stellar radiation field.
An X-ray counterpart to the Cocoon will trace the extended TeV emission if some of the TeV emission is in fact produced by inverse Compton scattering.

\citet{mizunoSuzakuObservationFermi2015} used {\it Suzaku} to constrain the X-ray 2--10 keV intensity from the Cocoon to less than $2.35\times 10^{-8}\rm~erg\,cm^{-2}\,s^{-1}\,sr^{-1}$ based on two on-source observations within the Cocoon and two off-source observations outside the Cocoon.
The {\it Suzaku} 2--10 keV intensity decreases monotonically with Galactic latitude, but the brightest source is actually labelled as one of their background targets (BG1).
The 2--10 keV upper limits are calculated by taking the two on-source observations (Source1 and Source2) and subtracting the fainter of the two backgrounds (BG2).
The {\it Suzaku} two off-source regions are within the TeV emitting region found by HAWC which was not known at the time, so they may include some signals.
The upper limit on X-ray flux from the entire Cocoon is found by extrapolating the upper limit intensity to the size of the Cocoon ($4.38\times10^{-3} \,\rm~sr$) finding an upper limit flux density of $6.41\times10^{-11}~\rm erg\,cm^{-2}\,s^{-1}$ at 1 keV assuming a spectral index equal to 2.

\begin{figure*}[ht]
\plotone{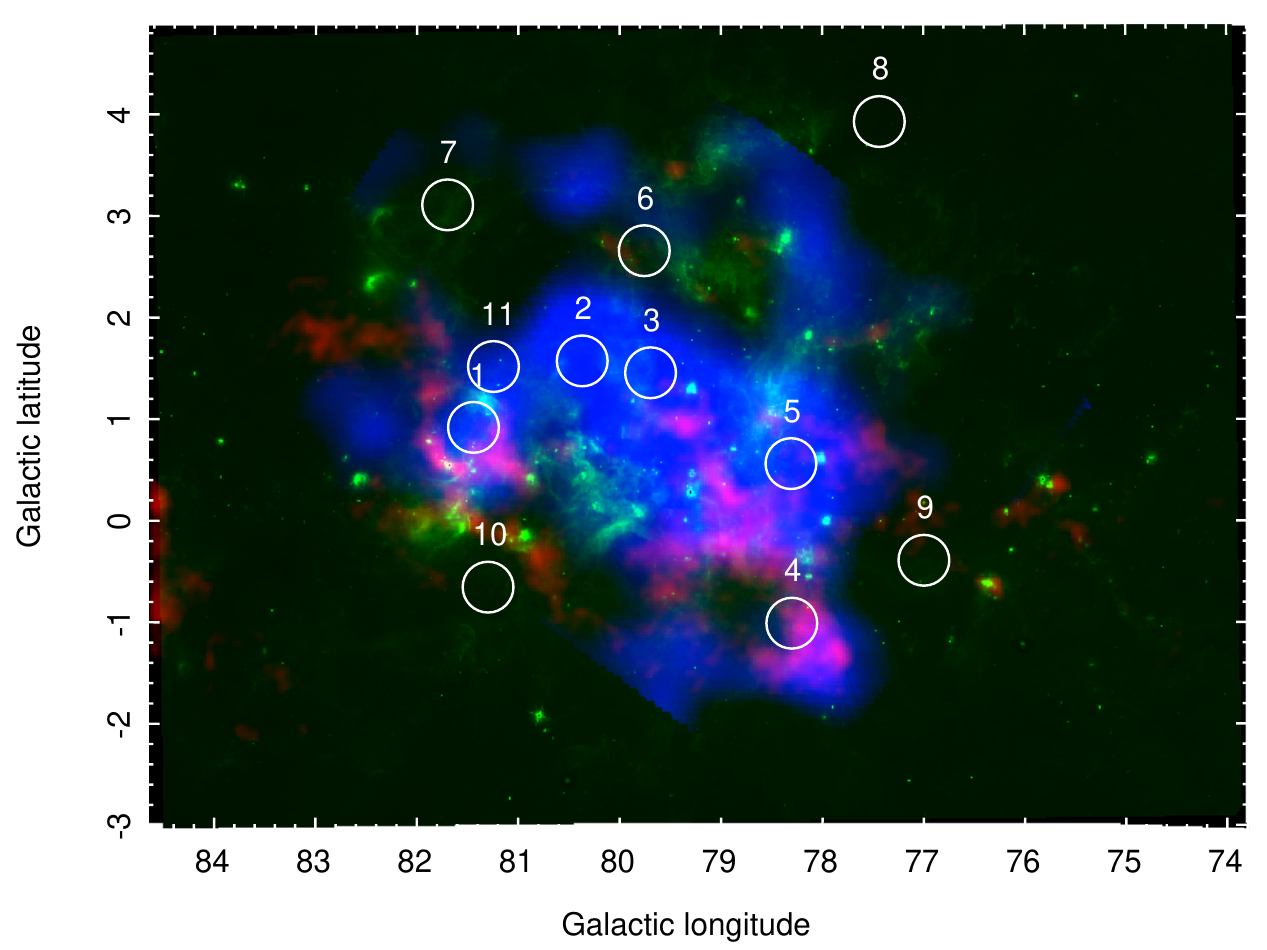}
\caption{Multi-wavelength map of the Cygnus Cocoon. Red is the Planck CO map \citep{planckcollaborationPlanck2013Results2014}, green is a WISE 12 micron mosaic \citep{wrightWIDEFIELDINFRAREDSURVEY2010}, and blue is the HAWC significance map above 1~TeV using 1,343 days of HAWC data \citep{{abeysekaraHAWCObservationsAcceleration2021}}. The blue cutoff   approximately corresponds to $3\sigma$. White circles indicate the 11 fields observed by \textit{Swift}-XRT and analyzed in this work.
\label{fig:sky_map}}
\end{figure*}

We observed eleven fields within and around the Cygnus Cocoon with the Neil Gehrels Swift Observatory's X-Ray Telescope ({\it Swift}-XRT) to search for evidence of leptonic emission.
We establish 2--10 keV upper limits on the intrinsic (unabsorbed) Cocoon intensity in 7 fields assuming an $E^{-2.5}$ spectrum as suggested by the leptonic model by \citet{abeysekaraHAWCObservationsAcceleration2021}.
Additionally, we derive upper limits on the solid angle integrated Cocoon emission assuming that the X-ray emission traces the TeV emission.

We describe our observations and data processing in Section \ref{sec:obsandproc}. Our spectral fitting is described in Section \ref{sec:spectral-fitting}. We discuss the implications for the leptonic TeV emission in Section \ref{sec:discussion}.


\begin{deluxetable}{ccccc}
\tablenum{1}
\tablecaption{Summary of Observations\label{tab:observations}.}

\tablehead{
\colhead{Field} & \colhead{Target ID} & \colhead{l [Deg]} & \colhead{b [Deg]} & \colhead{Exposure [s]} }
\startdata
  1 & 95932 & 81.418 & +0.921 & 10007\\
\ 2$^*$ & 95933 & 80.341 & +1.597 & 9817  \\
\ 3$^*$ & 95934 & 79.714 & +1.510 & 8256 \\
  4 & 95935 & 78.295 & -0.975 & 8309\\
  5 & 95936 & 78.327 & +0.586 & 9282\\
  6 & 95937 & 79.710 & +2.684 & 9610\\
\  7$^{\dagger}$ & 95938 & 81.683 & +3.110 & 9025\\
\  8$^{\dagger}$ & 95939 & 77.457 & +3.945 & 7554\\
\ \ 9$^{*\dagger}$ & 95940 & 76.973 & -0.402 & 9012 \\
\ \ 10$^{\dagger}$ & 95941 & 81.297 & -0.622 & 7994\\
\ 11 & 15110 & 81.246 & +1.521 & 10723\\
\enddata
\tablecomments{Dagger ($\dagger$) indicates that the field is outside of the Cocoon and used as background.
Star ($*$) indicates a field that is contaminated by stray light. The treatment of stray light contamination is discussed in Section \ref{sec:stray-light}. }
\end{deluxetable}

\section{Observations and Data Processing}
\label{sec:obsandproc}
\subsection{Observations}
\label{sec:observations}
{\it Swift} observed ten fields in or near the Cygnus Cocoon from May 27, 2021 to December 19, 2021, and one additional field as a target of opportunity from May 4, 2022 to June 13, 2022.
Each field was observed for approximately 10 ksec.
Seven targets (labeled 1-6, 11) are within the TeV emitting region and four (labeled 7-10) surround the Cocoon to provide a reference for the X-ray background.
The fields are shown in Figure \ref{fig:sky_map} with infrared, radio, and TeV observations of the Cocoon and summarized in Table \ref{tab:observations}.

\subsection{Data Processing}
\label{sec:processing}
We retrieved event lists and exposure maps from the HEASARC \citep{nasahighenergyastrophysicssciencearchiveresearchcenterheasarcHEAsoftUnifiedRelease2014}.
Each field was observed multiple times, so we produced an image by combining each of the event lists using {\it XSelect}.
We ran a point source detection algorithm in {\it XImage} to search for point sources with signal to noise greater than three.
For each point source found, we excluded a circle with 30$''$ radius around it from further processing.
We also excluded detector regions affected by stray light (see Section \ref{sec:stray-light}).
Fields 2, 3, and 9 were affected by stray light from nearby X-ray sources.
After stray light removal, Field 9 had too few counts to provide a useful constraint so we do not use it in the following analysis.
We extract a spectrum from the entire remaining detector after point source and stray light filtering.
We produced an exposure map for each field by summing the individual exposure maps.
We produced calibration files using the XRT tools with the extended source option enabled in the XRT {\tt xrtmkarf} task.


\begin{figure}[t!]
\centering
   \includegraphics[width=0.49\textwidth]{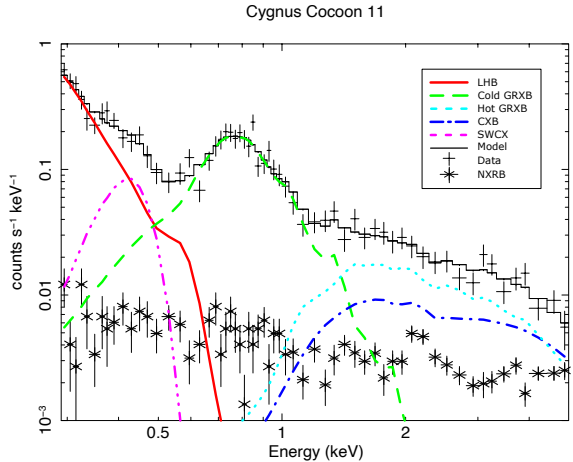}
\caption{Spectrum from Field 11. The data are shown with error bars overlaid with the overall model (black line). The model components are the Solar Wind Charge Exchange (SWCX; pink dot-dot-dot-dash), Local Hot Bubble (LHB; red solid), Cold Galactic Ridge X-Ray Background (GRXB; green dash), hot GRXB (cyan dot), and Cosmic X-Ray
Background (CXB; blue dot-dash). The non-X-ray background is plotted with error bars (black crosses) and dominates above 6 keV.
\label{fig:example_spectrum}}
\end{figure}

\subsection{Stray Light Filtering}
\label{sec:stray-light}
\textit{Swift}-XRT observations can be contaminated by stray light from X-ray sources outside the telescope field of view \citep{morettiNewMeasurementCosmic2009}.
Because we are searching for a source which is extracted from the entire detector, stray light contamination can strongly affect the spectrum.
In unprocessed images of Fields 2, 3 and 9, the stray light appears in a characteristic ring-like pattern.
The centers of Fields 2 and 3 are approximately 45$'$ from Cyg X-3 which is likely the source of the stray light which contaminates up to half of the detector for those observations.
The stray light in those fields is temporally variable with a hard power law spectrum consistent with a high mass X-ray binary like Cyg X-3.
The source of stray light in Field 9 is unknown, but it features a similar hard power law spectrum suggesting another X-ray binary.
We excluded contaminated detector regions in fields 2 and 3, and produced spectra from the remaining detector area.
As noted above, field 9 was too contaminated to produce a useful spectrum.
There are ten observations with useful data (1-8, 10, 11).
Below, ``all fields" refers to fields 1-8, 10, and 11. 

\section{Spectral Fitting}
\label{sec:spectral-fitting}

\subsection{Modelling}
\label{sec:model}
We include four components in the astrophysical background model: the Galactic Ridge X-Ray Background (GRXB), the Local Hot Bubble (LHB), Cosmic X-Ray Background (CXB), and the Solar Wind Charge Exchange (SWCX). The model for the Cocoon is an absorbed power law. These are summarized below:
\begin{enumerate}
    \item {\bf GRXB}: The GRXB is modeled with two absorbed Astrophysical Plasma Emission Code (APEC) \citep{smith2001collisional, foster2012updated} plasmas following previous studies of the GRXB \citep{mizunoSuzakuObservationFermi2015, 1986PASJ...38..121K}.
    We refer to these as the ``hot'' and ``cold'' GRXB.
    Both components are absorbed by the {\it tbabs} or {\it tbgrain} component.
    The column density of the cold GRXB is allowed to float in the fits while the hot GRXB absorption column density is frozen to the Galactic value (see Sec \ref{sec:column-density})

    \item {\bf LHB}: The LHB is modeled by an unabsorbed APEC plasma which dominates the spectra below 0.5 keV.

    \item {\bf CXB}: The CXB is modeled by an absorbed power law with normalization and spectral index frozen to the results of \citet{morettiNewMeasurementCosmic2009}.
    The absorption column density is frozen to the Galactic column density.

    \item {\bf SWCX}: We model the SWCX with a single line at 0.5 keV.
    In reality, the SWCX features many lines, but the {\it Swift}-XRT energy resolution at low energy is broad enough that the single line is adequate.

    \item {\bf Cocoon}: A hypothetical Cocoon component is modelled by an absorbed power law.
    Over the fit range 0.3--6.0 keV, a power law adequately represents the expected synchrotron emission.
    The spectral index was frozen to 2.5 motivated by the HAWC TeV result and the neutral hydrogen column density was frozen to the nearby column density (see Sec \ref{sec:column-density}). We also tested a spectral index of 2.0.
\end{enumerate}

We include a non-X-ray background spectrum using 43 ksec of data collected with the telescope sun shutter closed.
This spectrum and its response is included in {\it XSpec} as a background for all the spectra.
The non-X-ray background rate is approximately $2\times 10^{-8} \rm~counts\,s^{-1}\,keV^{-1}\,pixel^{-1}$.
Each of the observations include up to 38000 pixels (with 2.36'' pixel scale), which equates to a background rate $0.007 \rm~counts\,s^{-1}\,keV^{-1}$.
The spectra are dominated by non-X-ray background above 6 keV.
A fit for one field is shown in Figure \ref{fig:example_spectrum}.

\subsection{$N_\mathrm{H}$ Column Density}
\label{sec:column-density}

\begin{figure}
\plotone{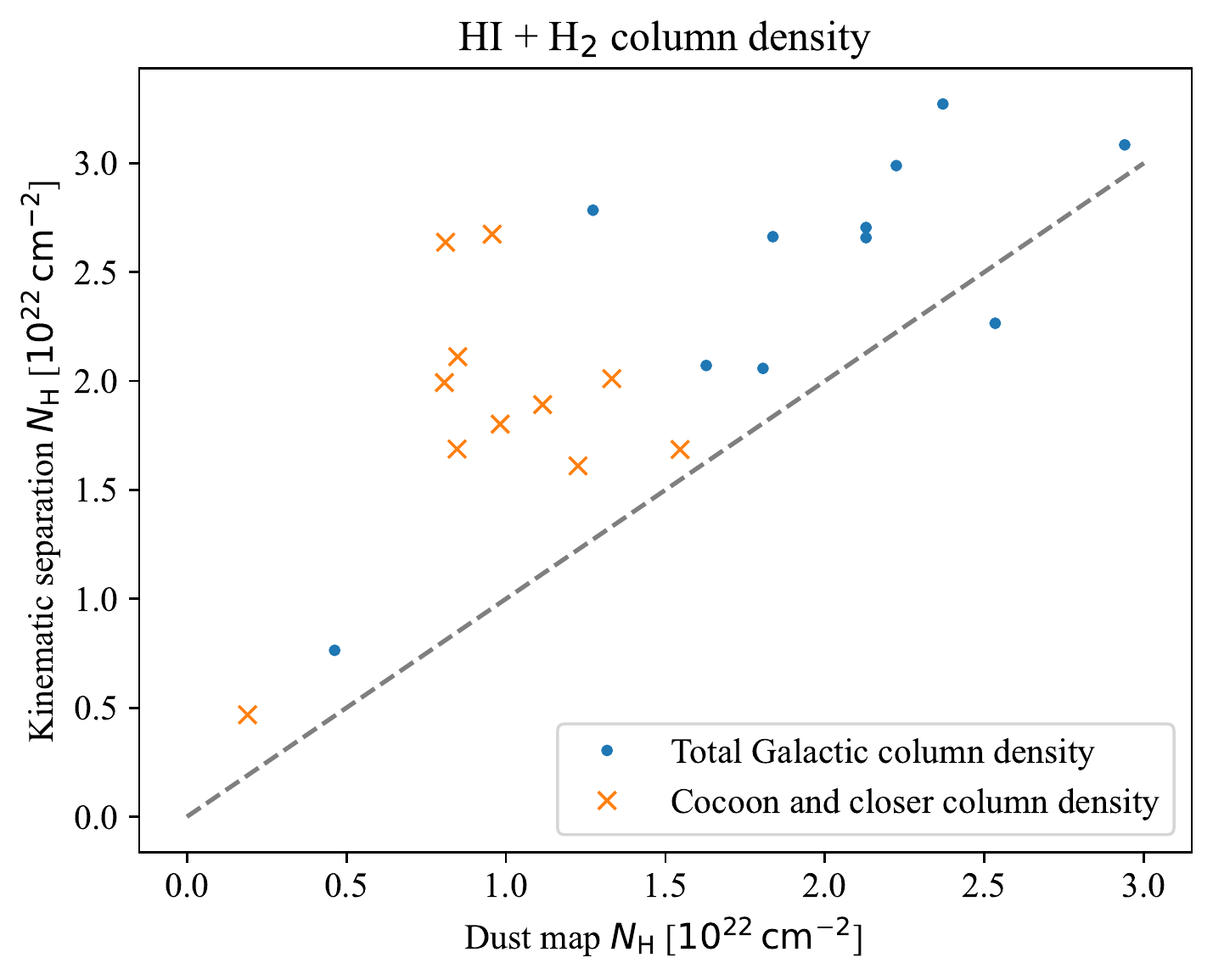}
\caption{Atomic and molecular hydrogen  column densities derived from visual extinction of stars and kinematic separation of 21-cm and CO lines combined with $\gamma$-ray data. The gray dashed line would be a perfect correspondence. The molecular hydrogen column density is the column density of hydrogen atoms that are in molecules.
\label{fig:nh}}
\end{figure}

The choice of column density, particularly the column density of nearby gas absorbing emission from the Cygnus region, affects the upper limit of the intrinsic emission from the Cocoon.
To quantify this impact we estimated the column density two ways.
First, we estimate the absorption column densities from 3-dimensional dust maps by \citet{green3DDustMap2019} using the \texttt{dustmaps} software package \citep{2018JOSS....3..695M}.
We extract the reddening, E(g-r), at 1.4 kpc and 10 kpc for each of our fields and convert it to E(B-V).
We convert E(B-V) to the neutral hydrogen column density using the relation from \citet{valencicINTERSTELLARDUSTPROPERTIES2015}.
Second, we use the column density maps derived by \citet{ackermannCosmicrayGasContent2012}.
The latter uses the Doppler shift of the 21-cm and CO emission lines to kinematically separate HI and H$_2$ into two regions: the Local Arm (where the Cygnus region and Cocoon are embedded), and beyond.
Additional visual extinction suggests additional dark neutral gas which is included in the nearby H$_2$ map. The conversion between CO emission and visual extinction to column density is based on $\gamma$-ray data.
The column densities are compared in Figure \ref{fig:nh}.
The average nearby kinematic separation column density is 2.1 times higher than the dust map result, and the average total Galactic column density is 1.4 times larger suggesting larger attenuation within the Cocoon.
\citet{wilmsAbsorptionXRaysInterstellar2000} suggest an H$_2$ fraction of 20\% which is assumed in the \texttt{tbabs} model;
however, the kinematic separation finds that the average fraction of hydrogen atoms in the molecular phase is approximately 50\% for this region.

\subsection{Hypothesis Testing}
\label{sec:hypothesis}

We performed a log-likelihood ratio test (LRT) for the models in the following sections where the null hypothesis is that there is no additional diffuse component and the test hypothesis is that there is additional emission as described in item~5 in Section~\ref{sec:model}.
Each model was fit by minimizing the C-statistic (equivalently maximizing the Poisson likelihood) with {\it XSpec} \citep{cashParameterEstimationAstronomy1979, 1996ASPC..101...17A} between 0.3--6.0 keV.
The models are nested, but the fit parameter space is constrained to keep the Cocoon normalization positive or zero.
The test statistic (twice the log-likelihood ratio) thus follows a ``$\chi^2/2$" distribution: $\rm~CDF(TS) =  {1}/{2}+\int_0^{TS} dx {\chi^2(x)}/{2} $ for $\rm~TS > 0$ \citep{mattoxLikelihoodAnalysisEGRET1996, protassovStatisticsHandleCare2002a}.
We verified that the test statistic does follow this distribution using Monte Carlo simulations where we generated synthetic data sets based on the best fit background model and performed a likelihood ratio test on each data set.
The resulting test statistic distribution had a $ {\chi^2}/{2}$ distribution with half the trials having $\rm~TS = 0$.
Intuitively, the unconstrained Cocoon normalization would be normally distributed around zero.
In the constrained case, the negative values are forced to be zero and thus have test statistic equal to zero.
For significance lower than $3\sigma$, upper limits for the parameters are defined by the 90\% confidence upper confidence interval on the parameter fit.

\subsection{Individual Fields}

\begin{figure}
\centering
   \includegraphics[width=0.49\textwidth]{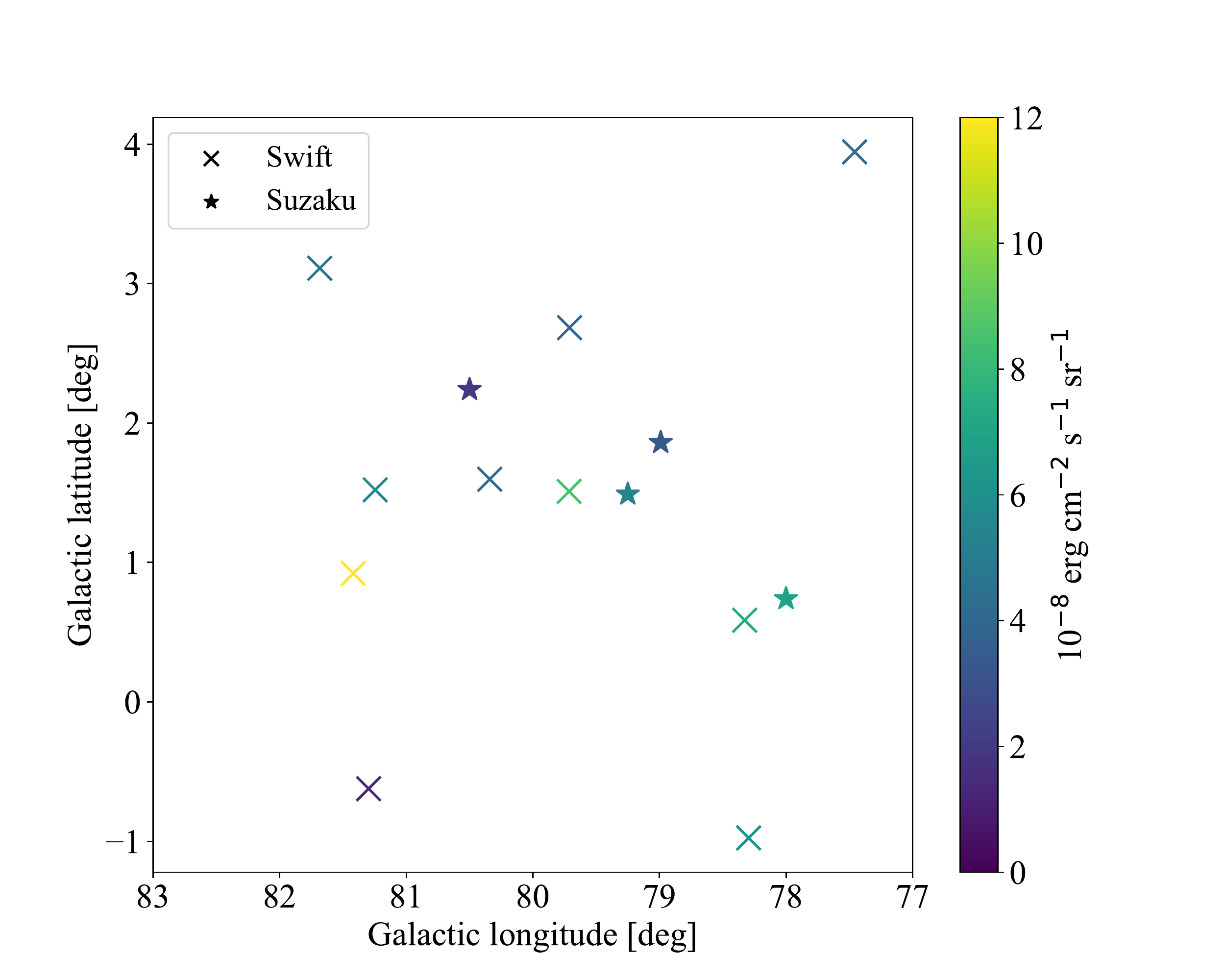}
\caption{Intensity for each field between 2--10 keV after subtracting CXB compared with the position of the center of each observation. Unlike \citet{mizunoSuzakuObservationFermi2015}, we do not find a monotonic relationship between Galactic latitude and hard X-ray emission.
\label{fig:intensity-latitude}}
\end{figure}

\input{individual_fit_table_single_v7.tex}

We fit each field with a background and a Cocoon plus background model.
The plasma temperatures, normalizations, and absorption column densities are allowed to vary in both fits.
In the latter case, the normalization of the Cocoon component is allowed to vary.
The best fit parameters for the background-only model are tabulated in Table \ref{tab:individual-fits}.
In general, the temperatures are similar except where the fit is poorly constrained by limited statistics at higher energy.
The background fields (7-10) in particular poorly constrain the hot GRXB.
The 2--10 keV intensity after subtracting CXB for each field is shown in Figure \ref{fig:intensity-latitude}.
Unlike \citet{mizunoSuzakuObservationFermi2015}, we find no monotonic relationship between the intensity and Galactic latitude.
Our Field 3 nearly overlaps the Suzaku Src 1 field and we find similar intensity in these two fields above 2 keV.
Field 3 and 6 are the nearest to Suzaku Src 2 but are 48$'$ and 65$'$ away.
The hard band intensity in Src 2 is similar to the intensity in Field 6.
We performed the hypothesis test described in \ref{sec:hypothesis} for an additional Cocoon component in the on-source fields modeled by an absorbed power law.
No field has a Cocoon component with significance greater than $3\sigma$.
The 90\% confidence upper limits on the Cocoon power law normalization are shown in Table \ref{tab:individual-fits}.
The mean upper limit is $2.0\times 10^{-8}\rm~erg\,cm^{-2}\,s^{-1}\,sr^{-1}$ between 2--10 keV.

\subsection{Integrated Limit on the Cygnus Cocoon}

\begin{figure*}
\plotone{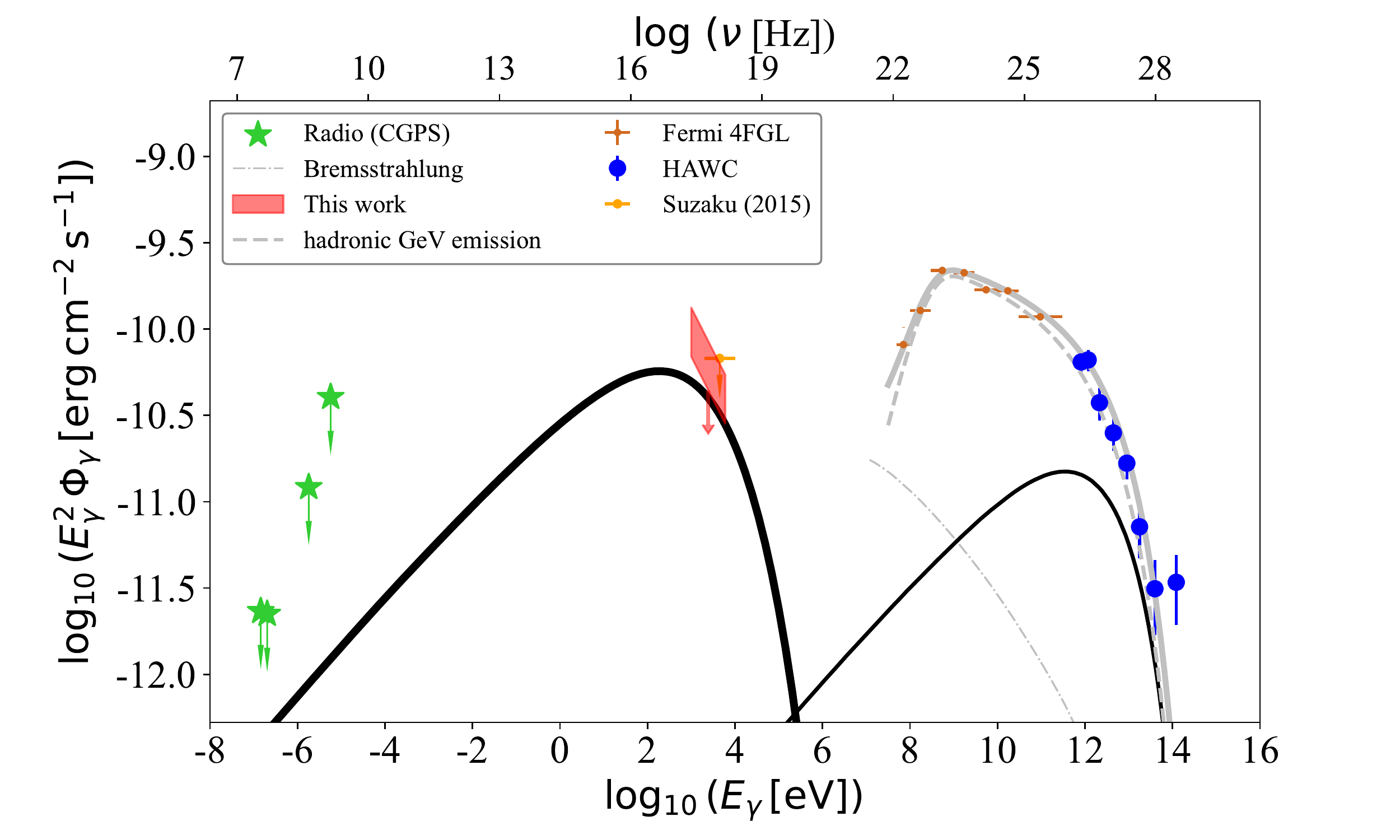}
\caption{Multi-wavelength spectral energy distribution with data and flux limits from HAWC \citep{abeysekaraHAWCObservationsAcceleration2021}, {\it Fermi}-LAT\citep{ackermannCocoonFreshlyAccelerated2011}, {\it Suzaku}-XIS \citep{mizunoSuzakuObservationFermi2015}, {\it Swift}-XRT, and CGPS \citep{2003AJ....125.3145T}. The red shaded region bounds the X-ray upper limits in this work for the two hydrogen column density models considered. The black curves represent the maximum synchrotron (thick black) and inverse Compton (thin black) emission allowed by our limit obtained using the kinematic separation column density. The injection electron spectrum is assumed to be $dN_e/dE_e\propto E_e^{-2}$ and magnetic field is set to 20~$\mu$G. The grey dash-dotted curve show the Bremsstrahlung emission by these electrons. The grey dashed curve indicates the pion decay emission by protons. The grey solid curve shows the sum of the emission components. Details of the background radiation model may be found in \cite{abeysekaraHAWCObservationsAcceleration2021}. 
\label{fig:sed}}
\end{figure*}

In addition to the individual fits, we jointly fit the spectra from all the fields assuming that a potential X-ray signal is proportional to the integrated 1-100 TeV surface brightness, $F_X\propto F_\gamma$. 
For the fields outside the Cocoon, $F_X$ of the Cocoon component is set to 0.
For the fields that have a non-zero HAWC flux, we tie the normalization of the Cocoon component to the Cocoon normalization in Field 1 multiplied by the ratio of the HAWC flux in each field divided by the HAWC flux in Field 1.
The model adds the constraint that the GRXB and LHB temperatures are the same across all the fields within the Cocoon region.
The 90\% confidence upper limit on the Cocoon normalization parameter is $6.7\times 10^{-4} \rm~cm^{-2}\,s^{-1}\,keV^{-1}$.
Multiplying this normalization parameter by the integrated TeV flux over the entire Cocoon (divided by the TeV flux in Field 1) and using the dust map based column density, we find the total flux 
\begin{equation}
    F = \int dE E \frac{dN}{dE\,dA\,dt} < 8.7\times 10^{-11}\,\rm erg\,cm^{-2}\,s^{-1} 
\end{equation}
between 2--10 keV when assuming $dN/dE \propto E^{-2.5}$.
This limit is shown in Figure \ref{fig:sed}.
Using a spectral index of 2 increased the 2--10 keV flux to $1.0~\times~10^{-10}~\rm erg\,cm^{-2}\,s^{-1}$.
The statistical significance of the likelihood ratio tests for $E^{-2.5}$ and $E^{-2}$ were $2.3\sigma$ and $3.4\sigma$, respectively.

We repeated the test using the column densities based on kinematic separation and calibrated on $\gamma$-ray data.
We also replaced the \texttt{tbabs} model with \texttt{tbgrain} which allows the user to set the fraction of hydrogen atoms in the molecular phase.
The significance of the log-likelihood ratio test reduces to $1.6\sigma$ and $1.7\sigma$ for $E^{-2.5}$ and $E^{-2.0}$ spectra respectively.
The corresponding 2--10~keV upper limits are $5.2\times 10^{-11}~\rm erg\,cm^{-2}\,s^{-1}$ and $4.4\times 10^{-11}~\rm erg\,cm^{-2}\,s^{-1}$
The results of both models are summarized in Table \ref{tab:joint-fit-parameters-free}.

\input{joint_fit_table.tex}

\section{Discussion}
\label{sec:discussion}

The X-ray upper limits derived from the {\it Swift}-XRT observations constrain the electron contribution to gamma rays above 1 TeV.
We adopt the leptonic model presented in \citet{abeysekaraHAWCObservationsAcceleration2021} which assumes that a population of relativistic electrons are continuously injected by the star clusters, diffuse in the Cocoon magnetic field, and inverse Compton scatter the stellar light fields and dust radiation field.
When fixing the magnetic field strength to 20 $\mu$G, which is estimated using the gas pressure based on infrared imaging \citep{ackermannCocoonFreshlyAccelerated2011}, the integrated {\it Swift}-XRT limit constrains the leptonic $\gamma$-ray emission at 1 TeV to $25\%$ of observed emission.

Alternatively, when freeing the magnetic field and allowing a dominant leptonic contribution above 10~TeV, the X-ray observation constrains the synchrotron radiation by the electrons and limits the field strength to $B\lesssim 10\,\mu$G. The maximum electron energy that can be accelerated by the stellar winds in this field is $E_{e,\rm max} \approx 34\,\eta^{1/2} (B/10\,\mu{\rm G})^{-1/2}(v_w/10^3\,\rm km\,s^{-1})\,\rm TeV$ \citep{blandfordParticleAccelerationAstrophysical1987, ackermannCocoonFreshlyAccelerated2011}, where $\eta = \delta B^2/B^2$ denotes the level of magnetic-field fluctuations and $v_w\sim10^3\,\rm km\,s^{-1}$ is the velocity of stellar winds. These electrons cannot explain the photons with energies above  $E_{e,\rm max}$. 

Although the spectrum in Field 1 cannot rule out the purely background hypothesis, it has significantly brighter hard emission than the rest of the field.
\citet{robertsASCACatalogPotential2001} searched for a X-ray counterparts to  GeV sources with ASCA including GeV J2035+4214 which lies within the Cygnus Cocoon.
Their observation of this source overlaps with Field 1 where they found three X-ray sources.
One of these sources (Src1) is clearly extragalactic because it is strongly absorbed and has a spectral index of 1.41.
This source was detected by {\it Swift}-XRT and excluded during data processing.
The other two sources (Src2 and Src3) have absorption column densities equal to 0.98 and 0.84 which suggest distances consistent with the Cygnus Cocoon.
These two sources have spectral indices equal to 2.02 and 2.44 and the latter is identified as an extended source.
In Chandra imaging of the same region, the extended source is in fact several confused point sources \citep{mukherjeeSearchPointSourceCounterpart2003}.
We did not detect Src2 in {\it Swift}-XRT images.
A comparison of spectra including and excluding Src3 made no apparent change in the shape of the {\it Swift}-XRT spectra.

\section{Summary}
We observed 11 fields in and near the Cygnus Cocoon with {\it Swift}-XRT to search for non-thermal X-ray emission.
The X-ray emission for each field is well described by a background model consisting of the Local Hot Bubble, Galactic Ridge X-Ray Background, and Cosmic X-Ray Background.
We establish upper limits for an additional power law component for each of the fields within the Cygnus Cocoon.
We also tested the hypothesis that non-thermal X-ray emission traces the TeV emission measured by HAWC and established an upper limit to this component. The limits thus obtained are sensitive to the hypotheses made to model X-ray absorption by interstellar matter, the impact of which is quantified.
The X-ray upper limits constrain the contribution of inverse Compton scattering to the Cocoon's TeV emission.
The X-ray limits can be explained by a magnetic field below equipartition within the Cocoon or by reducing the relativistic electron density.
In the latter case, the inverse Compton emission can explain up to one quarter of the TeV emission.

\begin{acknowledgments}
We thank Binita Hona for discussion on the HAWC significance map. The work of D.G. and K.F. is supported by the Office of the Vice Chancellor for Research and Graduate Education at the University of Wisconsin-Madison with funding from the Wisconsin Alumni Research Foundation. D.G. and K.F. acknowledge support from NASA (NNH20ZDA001N-Swift) and National Science Foundation (PHY-2110821). A.B. and K.L.P. acknowledge funding from the UK Space Agency.

\end{acknowledgments}

\bibliography{main.bib}{}
\bibliographystyle{aasjournal}

\end{document}

%% file: individual_fit_table_single_v7.tex
\begin{splitdeluxetable*}{ccc|CC|CD|DCCBDDD||CCC}
\decimals
\tablenum{2}
\tablecaption{Summary of Individual Background-Only Fit Results and Upper Limits\label{tab:individual-fits}}
\tablehead{
\colhead{Field} & \colhead{C-statistic} & \colhead{DoF} & \multicolumn2c{SWCX} & \multicolumn3c{LHB} & \multicolumn4c{Cold GRXB} & \multicolumn6c{Hot GRXB} & \multicolumn3c{Cocoon}\\
\multicolumn3c{} & \multicolumn1c{Energy} & \multicolumn1c{norm} & \multicolumn1c{$kT$} & \multicolumn2c{norm} & \multicolumn2c{$N_\mathrm{H}$} & \multicolumn1c{$kT$} & \multicolumn1c{norm} & \multicolumn2c{$N_\mathrm{H}$} & \multicolumn2c{$kT$} & \multicolumn2c{norm} & \multicolumn1c{$N_\mathrm{H}$} & \multicolumn1c{Spectral Index} & \multicolumn1c{norm} \\
\multicolumn3c{} & \multicolumn1c{keV} & \multicolumn1c{$\times 10^{-4}$} & \multicolumn1c{$\times 10^{-2}\rm~keV$} & \multicolumn2c{$\times 10^{-3}$} & \multicolumn2c{$\times 10^{22}\rm~cm^{-2}$} & \multicolumn1c{keV} & \multicolumn1c{$\times 10^{-3}$} & \multicolumn2c{$\times 10^{22} \rm~cm^{-2}$} & \multicolumn2c{keV} & \multicolumn2c{$\times 10^{-3}$} & \multicolumn1c{$\times 10^{22} \rm~cm^{-2}$} & \multicolumn1c{} & \multicolumn1c{$\times 10^{-4}$} \\
}
\startdata
1 & 464.96 & 479 & 0.69^{+0.05}_{-0.05} & 1.03^{+0.52}_{-0.50} & 7.26^{+0.79}_{-0.75} & 19.86^{+15.33}_{-7.52} & 1.33^{+0.18}_{-0.20} & 0.56^{+0.09}_{-0.13} & 9.81^{+5.32}_{-3.02} & 2.37^{\dagger} & 6.12^{+6.15}_{-2.32} & 7.48^{+1.92}_{-1.06} & 0.81^{\dagger} & 2.5^{\dagger} & <27.11\\
2 & 316.41 & 386 & 0.52^{+0.05}_{-0.04} & 1.21^{+1.05}_{-0.48} & 5.70^{+1.04}_{-0.98} & 62.35^{+134.45}_{-36.70} & 0.48^{+0.30}_{-0.23} & 0.30^{+0.12}_{-0.09} & 3.09^{+7.85}_{-1.99} & 1.84^{\dagger} & >11.10 & 2.48^{+0.49}_{-0.73} & 1.12^{\dagger} & 2.5^{\dagger} & <2.40\\
3 & 361.95 & 419 & 0.51^{+0.03}_{-0.03} & 1.63^{+1.05}_{-0.88} & 5.10^{+1.06}_{-1.13} & 99.10^{+390.24}_{-65.90} & 1.02^{+0.16}_{-0.24} & 0.21^{+0.07}_{-0.04} & 37.25^{+81.75}_{-27.66} & 2.22^{\dagger} & >6.69 & 4.41^{+1.36}_{-0.58} & 0.81^{\dagger} & 2.5^{\dagger} & <1.62\\
4 & 375.14 & 391 & 0.54^{+0.02}_{-0.02} & 6.63^{+1.70}_{-4.51} & 4.66^{+0.87}_{-0.75} & 284.31^{+661.94}_{-185.76} & 0.47^{+0.25}_{-0.19} & 0.39^{+0.12}_{-0.13} & 2.07^{+5.41}_{-1.00} & 2.94^{\dagger} & 4.58^{+11.08}_{-1.83} & 4.12^{+1.24}_{-0.99} & 0.96^{\dagger} & 2.5^{\dagger} & <9.67\\
5 & 438.13 & 455 & 0.45^{+0.02}_{-0.02} & 5.64^{+1.48}_{-2.54} & 2.19^{+0.55}_{-0.78} & 76255.50^{+600963.05}_{-75825.46} & 0.98^{+0.18}_{-0.14} & 0.15^{+0.03}_{-0.04} & 203.67^{+1116.13}_{-147.11} & 1.27^{\dagger} & 5.20^{+5.60}_{-1.69} & 4.46^{+0.70}_{-0.60} & 0.98^{\dagger} & 2.5^{\dagger} & <14.13\\
6 & 383.75 & 421 & 0.46^{+0.02}_{-0.02} & 7.62^{+5.55}_{-3.23} & 4.18^{+1.08}_{-2.26} & 587.26^{+5041.97}_{-475.13} & 0.40^{+0.11}_{-0.10} & 0.31^{+0.04}_{-0.04} & 12.25^{+8.73}_{-4.50} & 2.13^{\dagger} & 2.70^{+1.84}_{-0.80} & 4.01^{+1.01}_{-0.92} & 0.85^{\dagger} & 2.5^{\dagger} & <12.85\\
7 & 392.38 & 408 & 0.69^{+0.04}_{-0.10} & 1.79^{+1.34}_{-1.21} & 7.13^{+1.10}_{-0.79} & 21.33^{+26.16}_{-10.16} & 0.33^{+0.13}_{-0.15} & 0.61^{+0.08}_{-0.09} & 2.57^{+0.99}_{-0.78} & 1.63^{\dagger} & >3.53 & 2.42^{+0.84}_{-0.58} & 1.23^{\dagger} & 2.5^{\dagger} & -\\
8 & 339.51 & 401 & 0.51^{+0.02}_{-0.02} & 6.86^{+2.25}_{-2.25} & 5.49^{+1.16}_{-1.25} & 111.76^{+447.22}_{-70.08} & 0.48^{+0.09}_{-0.11} & 0.24^{+0.03}_{-0.02} & 42.83^{+28.97}_{-18.15} & 0.46^{\dagger} & 4.01^{+4.96}_{-1.49} & 2.81^{+0.61}_{-0.56} & 0.19^{\dagger} & 2.5^{\dagger} & - \\
10 & 358.83 & 385 & 0.73^{+0.03}_{-0.04} & 1.83^{+0.81}_{-0.82} & 7.58^{+0.78}_{-0.67} & 22.39^{+13.08}_{-8.01} & 1.45^{+0.32}_{-0.30} & 0.22^{+0.06}_{-0.05} & 90.14^{+317.51}_{-62.89} & 1.81^{\dagger} & 3.19^{+6.35}_{-1.26} & 3.16^{+1.23}_{-0.58} & 1.55^{\dagger} & 2.5^{\dagger} & - \\
11 & 400.99 & 447 & 0.45^{+0.02}_{-0.02} & 4.67^{+3.98}_{-2.53} & 5.43^{+0.94}_{-1.82} & 97.04^{+317.42}_{-52.89} & 0.20^{+0.14}_{-0.13} & 0.34^{+0.09}_{-0.05} & 2.86^{+2.46}_{-1.31} & 2.13^{\dagger} & 4.55^{+4.57}_{-1.45} & 3.80^{+0.71}_{-0.65} & 1.33^{\dagger} & 2.5^{\dagger} & <6.00 \\
\enddata
\tablecomments{A less-than sign or greater-than sign indicates an upper limit or lower limit with 90\% confidence. All errors are 90\% confidence intervals.}
\end{splitdeluxetable*}

%% file: joint_fit_table.tex
\begin{deluxetable*}{ccC|cCC}
\decimals
\tablenum{3}
\tablecaption{Integrated Cocoon Likelihood Ratio Test Results and Upper Limits\label{tab:joint-fit-parameters-free}}
\tablehead{
\multicolumn1c{C-statistic} & \multicolumn1c{DoF} & \multicolumn1c{Significance} & \multicolumn1c{$N_\mathrm{H}$ Model} & \multicolumn1c{Cocoon Spectral Index} & \multicolumn1c{2--10~keV Cocoon Flux}\\
\multicolumn5c{} & \multicolumn1c{$\times 10^{-11}\rm~erg\,cm^{-2}\,s^{-1}$}\\
}
\startdata
\hline
3946 & 4217 & 2.3\sigma & Dust map & 2.5 & <8.7\\
3946 & 4217 & 3.4\sigma & Dust map & 2.0  & <10.0\\
3995 & 4217 & 1.6\sigma & Kinematic separation & 2.5  & <5.2\\
3995 & 4217 & 1.7\sigma & Kinematic separation & 2.0  & <4.4\\
\enddata
\tablecomments{The C-statistic is for the background-only fit. The $N_\mathrm{H}$ models are described in Sec \ref{sec:column-density}.}
\end{deluxetable*}